\documentclass[amsmath,preprintnumbers]{revtex4}
\usepackage{graphicx}
\usepackage{bm}
\newcommand{\pd}[2]{\frac{\partial {#1}}{\partial {#2}}} 

\newcommand{\al}{{\alpha}}
\newcommand{\be}{{\beta}}
\newcommand{\ga}{{\gamma}}

\newcommand{\de}{{\delta}}
\newcommand{\De}{{\Delta}}

\newcommand{\ka}{{\kappa}}
\newcommand{\la}{{\lambda}}

\newcommand{\ta}{{\tau}}
\newcommand{\om}{{\omega}}
\newcommand{\Om}{{\Omega}}

\begin{document}

\title{Studies of group velocity reduction and pulse regeneration with 
and without the adiabatic approximation}

\author{M.G. Payne$^{1}$, L. Deng$^{2}$, Chris Schmitt$^{3}$, and Shannon Anderson$^{3}$}
\affiliation{$^{1}$ Department of Physics, P.O. Box 8031, Statesboro, GA30460}
\affiliation{$^{2}$ Electron and Optical Physics Division, NIST, Gaithersburg, MD 20899-8410 }
\affiliation{$^{3}$ McNair Scholar in Physics, Georgia Southern University, Statesboro, GA 30460 }

\begin{abstract}

We present a detailed semiclassical study of the propagation of a pair of optical fields in
resonant media with and without adiabatic approximation.  In the case of near
and on resonance excitation, we show detailed calculations, both analytically
and numerically, on the extremely
slowly propagating probe pulse and the subsequent regeneration of a pulse
via a second pulse from the coupling laser.  Further discussions on the adiabatic approximation provide
many subtle details of the processes, including the effect on the pulse
width of the regenerated optical field.  Indeed, all features of the optical
pulse regeneration and most of the intricate details of the process
can be obtained with the present treatment without invoking a full field
theoretical method. For very far off resonance excitation, we show that the
analytical solution is nearly detuning independent, a surprising result that is
vigorously tested and compared to numerical calculations with very good
agreement.

\end{abstract}

\maketitle

\section{Introduction}

Propagation of optical pulses in a resonant medium has long been an active field
of research.  The advances
of laser technologies have stimulated further development and understanding of
the propagation of optical pulses under various conditions, in particular in the
field of ultra-short and extremely high intensity optical pulse generation and
propagation under adiabatic conditions.  Early works in this field include
the interpretation of the nonlinear wave mixing in terms of an adiabatic
following model [1], counter-intuitive pulse sequence [2] and induced
transparency [3].  Later, Eberly and co-workers [4] developed a theory for the propagation
of optical field under the condition that atomic response can be treated
adiabatically.  Further development on the propagation properties of a pulse pair
under the similar adiabatical conditions resulted in substantial understanding of
the evolution of optical pulses in a resonant media.  In particular, Grobe and
co-workers first proposed the creation and recall of excitations distributed
in media [5].  Recently, the field has experience a renewed interest and resurgence following the
demonstration of extremely slow propagation of optical pulse both in
ultra-cold and high temperature atomic gases [6].  Experimental studies on
possible excitation of the atomic spin wave excitations with
ultra-slow optical field, and the subsequent ``revival" of an optical field with
frequencies near that of the ultra-slow field have promised future applications
in information technology [7-9].  Theoretical works on such spin wave excitations and
the subsequent ``revival" of an optical filed have also been formulated in a ``dark-state polariton" theory [10],
which basically is a field theoretical reformulation of the adiabatic approximation of pulse pair propagation in
a resonant medium originally formulated before [4,5].  These new treatments have
provided some further understanding of the process, especially from the view
point of atomic spin wave excitation and their relation to atomic coherence.
\vskip 5pt
Here, we present a study on the propagation and ``retrieval" of ultra-slow optical
fields.  We investigate both analytically and numerically the propagation
problem, with and without adiabatic approximation, for the cases of near resonance, on-resonance,
and far detuned from resonance.  In all cases, we show that the slow group velocity propagation
and pulse ``revival" is independent of the one-photon detuning by the probe laser,
providing that the probe and coupling lasers are tuned to exact two-photon
resonance and certain conditions are satisfied.
That is, the results on off-resonance excitations are identical to the adiabatic theory formulation
for the exact on-resonance case treated in [4].  The main contribution
of our work, however, is the detailed analysis, both analytical and
numerical, on the extremely slow propagation of the optical field, the
corresponding atomic response, and the ``revival" of an optical field when a
second time-delayed coupling pulse is injected into the medium.  These
results, to the best of our knowledge,
are not contained in the original studies by Eberly and co-workers and all the
subsequent studies, including all
recent theoretical works, on the subject.  In addition, we show that all the experimental
observables can be well predicted with the usual treatment combining
the classical electrodynamics  and the three-state model, without invoking a full field
theoretical methodology. This is not surprising, since when a detailed handling of spontaneous
emission is not critical and the fields are not extremely weak, a mean-field approximation
to the quantum treatment of the electromagnetic field is valid.
In Section II, we first present an adiabatic theory
for the case where non-vanishing one-photon detuning exists.  This treatment
is very similar to the original treatment given in [4] except that the latter
deals exclusively with the case of on resonance excitation.  In Section III, we
examine the propagation of a pulse pair and give two examples on how to
obtain the evolution of the optical fields.  Section IV constitutes the main
contribution of the present study.  We discuss in detail the propagation
of the pulse pair in the adiabatic limit and deduce analytically substantial
insight and subtle understanding on the slow propagation of the probe field,
the optical field-atomic spin cohenrence conversion, and the characteristics
of the ``retrieved" optical field.  These results are not available in any of
previous studies, and we believe they are very useful in providing a complete
picture for the physics involved.  In Section V, we compare predictions and
estimates derived in Section IV with numerical simulations.  Section VI is devoted
to the case where the probe laser is tuned far from resonance, i.e. a Raman
excitation scheme is
employed.  In this case, a different set of conditions are required for the adiabatic
theory to be accurate. We show, however, that when certain conditions are satisfied, the
solution is independent of the detuning, a surprising result that does not, to
our best knowledge, exist in literature.  Finally, the results are compared with
numerical calculations for vigorous testing.  In Section
VII, we present a conclusion for our study.

\section{ Preliminary}

In this section, we examine the propagation of a pair of optical pulse
in resonant media under the conditions where the adiabatic approximation can
be made.  We present here an expanded investigation based on the early
treatment given in [4].  The difference is the inclusion of the non-vanishing
one-photon detuning since the original treatment assumed
exact on one-photon resonance excitations.  We will show that
this non-vanishing one-photon
detuning leads to correct predictions about pulse propagation, coherent population
transfer and storage, and the regeneration of an optical field when a time delayed second pulse
is provided from the coupling laser.  It should be pointed out that all theoretical
studies published recently on this subject also assume exact on resonance
tuning by the probe laser as originally assumed in Ref.[4].  Readers should consult a series
pioneer studies on the on-resonance case given in [4].
\vskip 10pt
Consider a $\Lambda$\ system as depicted in Figure 1.  We assume that a probe laser
($E_p$, frequency $\om_p$)
is tuned on or near resonance with the $|1>\rightarrow |2>$\ transition, and
a coupling laser ($E_c$, frequency $\om_c$) is tuned so that exact two-photon resonance
between states $|1>$ and $|3>$ is achieved.  Assume atomic wave function of the form
\begin{equation}
|\Psi(z,t_r)>=a_1 e^{-i\om_1t}|1>+a_2 e^{-i\om_2t}|2>+a_3 e^{-i\om_3t}|3>,
\end{equation}
we thus find the following atomic equations of motion
\begin{subequations}
\begin{eqnarray}
\pd{a_1}{t_r}&=&i\Om_{p}e^{-ik_pz}a_2,\\
\pd{a_2}{t_r}&=&i\Om_{p}^{*}e^{ik_pz}a_1+i\Om_{c}^{*}e^{ik_cz}a_3+i\left(\delta+i\frac{\ga_2}{2}\right)a_2,\\
\pd{a_3}{t_r}&=&i\Om_{c}e^{-ik_cz}a_2, 
\end{eqnarray}
\end{subequations}
where $a_j$ and $\ga_j$ are the $jth$ amplitude of the atomic wave function and decay
rate, respectively.  $\Om_p$\ and $\Om_c$\ are half-Rabi frequencies, given by
$\Om_p^{*}=D_{21}E_p/(2\hbar)$, $\Om_c^{*}=D_{23}E_c/(2\hbar)$ with $D_{ij}$ being
the dipole moment of the relevant transition.  Also, $\om_p=\om_2-\om_1+\delta$\,
$\om_c=\om_2-\om_3$, and the quantity $t_r=t-z/c$\ is the retarded time which is
the combination of $t$ and $z$ that the laser field amplitudes depend on in vacuum.
\vskip 5pt
In order to correctly describe the propagation of the optical pulse, the atomic
equations of motion (2) must be simultaneously solved self-consistently
with  Maxwell equations.   In the limit
of plane waves and slowly varying amplitudes, the positive frequency part of
these fields satisfy
\begin{subequations}
\begin{eqnarray}
\left(\pd{\Om_{p}^{*}}{z}\right)_{t_r}&=&i\kappa_{12}\tau A_1^* A_2, \\
\left(\pd{\Om_{c}^{*}}{z}\right)_{t_r}&=&i\kappa_{32}\tau A_3^* A_2, 
\end{eqnarray}
\end{subequations}
where the amplitudes $A_1$, $A_2$, and $A_3$\ differ by position dependent
phase factors from $a_1$, $a_2$, and $a_3$, and we have introduced notations $\kappa_{12}=2{\pi}N\om_p|D_{12}|^2/(\hbar c)$\
and $\kappa_{32}=2{\pi}N\om_c|D_{32}|^2/(\hbar c)$\ with $N$ being the concentration.
\vskip 5pt
In order to evaluate the polarization terms on the right hand sides of Eq.(3a,3b),
it is necessary to obtain the amplitudes of the atomic wavefunction.  The problem
is intrinsically difficult because both Eq.(2) and (3) must be solved simultaneously.
A great simplification can be achieved if one invokes the adiabatic approximation
while evaluating the atomic wavefunction.  The wavefunction obtained with the adiabatic
limit will then be used to calculate the optical field in a self-consistent manner.
This is the spirit of all adiabatic approximation based optical propagation
problem and will be the method used in the present work.

\vskip 5pt
In an adiabatic treatment we introduce $A_1=a_1$, $A_2=e^{ik_pz}a_2$, and $A_3=e^{ik_cz}a_3$\
in order to remove the $z$\ dependent phase factors. The adiabatic eigenvalues
of the equations of motion can then be obtained from the characteristic
equation given by

$$
\la^3-\left(\delta+i\frac{\ga_2}{2}\right)\la^2-(|\Om_{p}|^2+|\Om_{c}|^2)\la=0.
$$
They are
\begin{subequations}
\begin{eqnarray}
\la_0&=0, \\
\la_{+}&=&\left(\frac{\delta+i\frac{\ga_2}{2}}{2}\right)+\sqrt{|\Om_p|^2+|\Om_c|^2+\left(\frac{\delta+i\frac{\ga_2}{2}}{2}\right)^2},\\
\la_{-}&=&\left(\frac{\delta+i\frac{\ga_2}{2}}{2}\right)-\sqrt{|\Om_p|^2+|\Om_c|^2+\left(\frac{\delta+i\frac{\ga_2}{2}}{2}\right)^2}. 
\end{eqnarray}
\end{subequations}
As can be seen, the inclusion of the non-vanishing one-photon detuning
changes the eigenvalues. For sufficiently small $\delta$, this condition
reduces correctly to that given in [4].
\vskip 5pt
The adiabatic condition requires that the eigenvalue $\la_0\tau$\ always
differ from $\la_{\pm}\tau$\ by an amount that is large compared with unity.
When $|\delta\tau|>>|\Om_{c}|$\ and $|\de\tau|>>|\Om_{p}|$, however, additional
conditions are required.  In fact, we require that the pulse length of the
coupling laser be much longer than that of the probe laser, and $|\Om_{c}\tau|^2/|\de\tau|>>1$
be satisfied.  Therefore, during the entire pulse of the
probe laser, the two-photon transition between $|1>$\ and $|3>$\ is shifted
well outside the bandwidth of the probe laser.  In another words, the overlap
of the probe pulse band width with the resonance must be avoided.
\vskip 5pt
We now focus on the eigenvector corresponding to $\la_0=0$, since it is this
adiabatic state that corresponds to the population all being in state $|1>$\
before the laser pulses.  In the lowest order of the adiabatic approximation
Eqs.(2a) and (2c) imply that $a_2=0$. Eq.(2b) then yields
\begin{equation}
A_1=-\frac{\Om_{c}^{*}}{\Om_{p}^{*}}A_3. 
\end{equation}
Since, for a closed system, $|a_1|^2+|a_2|^2+|a_3|^2=1$\  we can write
\begin{subequations}
\begin{eqnarray}
A_1(z,t_r)&=&{\frac{\Om_c^{*}(z,t_r)}{\sqrt{|\Om_c(z,t_r)|^2+|\Om_p(z,t_r)|^2}}},\\
A_2(z,t_r)&=&{0}, \\
A_3(z,t_r)&=&{-\frac{\Om_p^{*}(z,t_r)}{\sqrt{|\Om_c(z,t_r)|^2+|\Om_p(z,t_r)|^2}}}.
\end{eqnarray}
\end{subequations}
It is important to realize that solutions (6a-6c) are not sufficiently accurate
for many propagation problems.  This is because that
even though numerical solutions show that $A_2$\ is much smaller than the amplitudes
of the other states, it cannot be exactly zero without assuming the laser
pulses are propagating at exactly the vacuum speed of light.  In many situations,
however, this assumption of the speed of laser pulses is not valid.  This is particularly
true in the case of sufficiently high optical density and very near a strong
resonance excitation where significant modification of the propagation velocity is
anticipated.  In order to accurately account the propagation effect in the regime
where significant dispersion effect is encountered,  one
must seek the next order correction to the lowest order approximation where the
small quantities such as $A_2$ becomes critically important.  To do this, one
can use either Eq.(2a) or Eq.(2c) to
derive a second approximation for $A_2$, thereby obtains the first order
correction to the lowest order adiabatic approximation.  Substitute Eq.(6a) into
Eq.(2a) or Eq.(6c) into Eq.(2c) we obtain
\begin{equation}
A_2(z,t_r)=-\frac{i}{\Om_p}\pd{A_1}{t_r}=-\frac{i}{\Om_p}\frac{\partial}{\partial t_r}\left(\frac{\Om_c^{*}}{\sqrt{|\Om_c|^2+|\Om_p|^2}}\right),\tag{7a}
\end{equation}
and
\begin{equation}
A_2(z,t_r)=-\frac{i}{\Om_c}\pd{A_3}{t_r}=\frac{i}{\Om_c}\frac{\partial}{\partial t_r}\left(\frac{\Om_p^{*}}{\sqrt{|\Om_c|^2+|\Om_p|^2}}\right).\tag{7b}
\end{equation}
Within the adiabatic approximation and the slowly varying amplitude approximation
one can show that $\Om_p(z,t_r)$\ and $\Om_c(z,t_r)$\ can be chosen to be real.
It is then trivial to show that two forms for $A_2$\ as shown in Eqs.(7a,7b)
are self consistent.  It is interesting to note that even though having $\delta$\ non-zero makes the criteria
for the validity of the adiabatic approximation different, it does not change
the solution for the state amplitudes.
\vskip 5pt
As usual with the adiabatic approximation used in a resonance situation, we must
have $\Om_c$\ already strong when $\Om_p$\ starts to build up.  This means that
$\Om_c\tau>>1$, where $\tau$\ is a measure of the time over which
$\Om_c$, or $\Om_p$\ change significantly.  One scenario where
the adiabatic approximation would hold through the whole laser pulse is
if the two lasers peak at the same time, but the pulse length of the coupling laser
is much longer. This feature of the pulse lengths and the requirement $|\Om_c\tau|>>1$\ through
the pulse length of the probe laser will make results based on the adiabatic
approximation quite accurate.  In essence, one must keep both $|\la_{+}\tau|>>1$\
and $|\la_{-}\tau|>>1$\ throughout the pulse of the probe laser, e.g. there
is a wide gap between $\la_0=0$\ and the other eigenvalues so that curve crossing
is avoided throughout the entire pulse length.  We will now investigate
the use of the adiabatic approximation in the propagation problem.
\begin{subequations}
\begin{eqnarray}
&& \nonumber
\end{eqnarray}
\end{subequations}

\section{The propagation problem}

In order to treat the propagation of the laser pulses through a resonant medium,
the polarization of the medium as a function of position and time must be obtained.
Using the atomic wave function obtained in Sec. II, we get
\begin{eqnarray}
P(t)&=&N<\Psi(z,t_r)|\hat D|\Psi(z,t)>\nonumber \\
&=&A_1^{*}(z,t_r)A_{2}(z,t_r)e^{ik_pz-i\om_pt}ND_{12}+A_3^{*}(z,t_r)A_2(z,t_r)e^{ik_cz-i\om_ct}ND_{32}+c.c.
\end{eqnarray}
where $N$ is the concentration. For the positive frequency parts of the
polarizations at the probe and coupling laser frequencies, we have
\begin{subequations}
\begin{eqnarray}
P_{\om_p}^{+}&=&ND_{12}e^{ik_p-i\om_pt}A_1^{*}A_2, \\
P_{\om_c}^{+}&=&ND_{32}e^{ik_c-i\om_ct}A_3^{*}A_2. 
\end{eqnarray}
\end{subequations}
Using the first order correction to $A_2$, i.e. Eqs.(7a,7b), we find, for the amplitude of $P_{\om}^{+}$\
\begin{subequations}
\begin{eqnarray}
P_{\om_p,0}&=&iND_{12}\frac{1}{\sqrt{|\Om_c|^2+|\Om_p|^2}}\frac{\partial}{\partial t_r}\left(\frac{\Om_p^{*}}{\sqrt{|\Om_c|^2+|\Om_p|^2}}\right),\\
P_{\om_c,0}&=&iND_{32}\frac{1}{\sqrt{|\Om_c|^2+|\Om_p|^2}}\frac{\partial}{\partial t_r}\left(\frac{\Om_c^{*}}{\sqrt{|\Om_c|^2+|\Om_p|^2}}\right).  
\end{eqnarray}
\end{subequations}
Substitute these polarizations into Eq.(3a,3b), we obtain
\begin{subequations}
\begin{eqnarray}
\left(\pd{\Om_{p}^{*}}{z}\right)_{t_r}&=&-\frac{\ka_{12}\tau}{\sqrt{|\tau\Om_c|^2+|\tau\Om_p|^2}}\frac{\partial}{\partial t_r/\tau}\left(\frac{\tau\Om_p^{*}}{\sqrt{|\tau\Om_c|^2+|\tau\Om_p|^2}}\right),\\
\left(\pd{\Om_{c}^{*}}{z}\right)_{t_r}&=&-\frac{\ka_{32}\tau}{\sqrt{|\tau\Om_c|^2+|\tau\Om_p|^2}}\frac{\partial}{\partial t_r/\tau}\left(\frac{\tau\Om_c^{*}}{\sqrt{|\tau\Om_c|^2+|\tau\Om_p|^2}}\right).  
\end{eqnarray}
\end{subequations}
\vskip 5pt
Equations (11) are the key equations describing propagation of a pulse pair
within the adiabatic approximation.
By using the adiabatic approximation for the atomic
response one needs to solve only two nonlinear partial differential
equations, instead of solving five nonlinear partial differential equations
Eq.(2a-2c) and Eq.(3a,3b).
\vskip 5pt
To further illustrate the effectiveness of the adiabatic approximation, we now
examine two special cases.
\vskip 5pt
We proceed by first noticing that the quantity
\begin{equation}
F=\frac{|\Om_p|^2}{\ka_{12}}+\frac{|\Om_c|^2}{\ka_{32}},
\end{equation}
represents the sum of the photon fluxes at $\om_p$\ and $\om_c$\ divided by
the concentration of the medium through which the waves propagate. Differentiate
$F$\ with respect to $z$ while holding $t_r$\ fixed and make use of Eq.(11), we obtain
\begin{equation}
\frac{\partial}{\partial z}\left(\frac{|\Om_p|^2}{\ka_{12}}+\frac{|\Om_c|^2}{\ka_{32}}\right)=0.
\end{equation}
Thus, $F$\ does not depends explicitly on $z$.  This
permits us to evaluate $F$\ by evaluating it at $z=0$, the entrance to the atomic
vapor cell, e.g.
\begin{equation}
F(z,t)=F(z=0,t)=\frac{|\Om_p(0,t)|^2}{\ka_{12}}+\frac{|\Om_c(0,t)|^2}{\ka_{32}}.
\end{equation}
Thus, whenever $|\Om_c(z,t_r)|^2/\ka_{32}+|\Om_p(z,t_r)|^2/\ka_{12}$\ occurs
it can be replaced by $F(t_r)$, as determined in Eq.(14).  An exact relation
for this quantity can also be directly derived from Eq.(2a-2c) and Eq.(3a,3b).
One obtains
\begin{equation}
\frac{\partial}{\partial z}\left(\frac{|\Om_p|^2}{\ka_{12}}+\frac{|\Om_c|^2}{\ka_{32}}\right)=-\pd{|A_2|^2}{t_r}-\ga_2|A_2|^2,\nonumber
\end{equation}
This relation indicates that anytime $|A_2|^2$\ is very small and slowly
varying, and the decay rate of this state is not too fast, this total photon
flux is close to depending only on $t_r$.  The lack of dependency of this
quantity on $z$ when the full set of equations is solved numerically is an important
test for the validity of the adiabatic approximation.

\subsection*{Case 1. $\ka_{12}=\ka_{32}$}

In this situation the quantity $|\Om_p(z,t_r)|^2+|\Om_c(z,t_r)|^2=\ka_{12}F(t_r)$.
We now use this fact in Eqs.(11a,b) to obtain an analytical approximation
that can be compared with numerical solutions to Eqs.(2a-2c) and (11a,b).
\vskip 5pt
Since $\sqrt{|\Om_c|^2+|\Om_p|^2}$\ does not depend on $z$ when $t_r$\ is held
fixed, dividing both sides of Eqs.(11a,b) by this quantity yields
\begin{subequations}
\begin{eqnarray}
\frac{\partial}{\partial z}\left(\frac{\Om_p^{*}\tau}{\sqrt{|\Om_c\tau|^2+|\Om_p\tau|^2}}\right)&=&-\frac{\ka_{12}\tau}{|\Om_c\tau|^2+|\Om_p\tau|^2}\frac{\partial}{\partial t_r/\tau}\left(\frac{\Om_p^{*}\tau}{\sqrt{|\Om_c\tau|^2+|\Om_p\tau|^2}}\right),\\
\frac{\partial}{\partial z}\left(\frac{\Om_c^{*}\tau}{\sqrt{|\Om_c\tau|^2+|\Om_p\tau|^2}}\right)&=&-\frac{\ka_{12}\tau}{|\Om_c\tau|^2+|\Om_p\tau|^2}\frac{\partial}{\partial t_r/\tau}\left(\frac{\Om_c^{*}\tau}{\sqrt{|\Om_c\tau|^2+|\Om_p\tau|^2}}\right).
\end{eqnarray}
\end{subequations}
A few substitutions make an analytical solution to these equations relatively
obvious.  Let
\begin{subequations}
\begin{eqnarray}
W_p&=&\frac{\Om_p^{*}\tau}{\sqrt{|\Om_c\tau|^2+|\Om_p\tau|^2}}, \\
W_c&=&\frac{\Om_c^{*}\tau}{\sqrt{|\Om_c\tau|^2+|\Om_p\tau|^2}}, \\
v(t_r)&=&\int_{-\infty}^{t_r/\tau}\left(|\Om_{c}(0,t_r^{'})\tau|^{2}+|\Om_{p}(0,t_r^{'})\tau|^{2}\right)d\left(\frac{t_r^{'}}{\tau}\right),\\
u(z)&=&\int_{0}^z\ka_{12}\tau dz^{'}. 
\end{eqnarray}
\end{subequations}
In terms of these dimensionless quantities Eqs.(15) become
\begin{subequations}
\begin{eqnarray}
\pd{W_p}{u}+\pd{W_p}{v}&=&0, \\
\pd{W_c}{u}+\pd{W_c}{v}&=&0, 
\end{eqnarray}
\end{subequations}
where general ``travelling wave" type solutions are immediately obtained as
\begin{subequations}
\begin{eqnarray}
{W_p}&=&F_p(v-u), \\
{W_c}&=&F_c(v-u).
\end{eqnarray}
\end{subequations}
This special case provides a very useful test case for examining the validity
of the adiabatic approximation for the ``trapped" light problem.  The functions
$F_p$\ and $F_c$\ are easily determined by evaluating at $z=0$ (therefore, $u=0$) and using
the fact that the laser fields are known as a function of $t$\ and hence $v$.
When the second coupling laser pulse is sent into the medium after a time
delay, the predictions about
the ``revival" of the probe laser are all contained in this solution.  The
tabulation of $F_p$\ remains the same as long as only a second coupling laser
is sent into the medium after some delay time.

\subsection*{Case 2. $|\Om_{p}(0,t)\tau|<<|\Om_{c}(0,t)\tau|$ with $\Om_p\tau<<1$}

In this limit there can never be a significant population in state $|3>$.  Therefore,
to a good approximation we have $\Om_{c}(z,t_r)=\Om_{c}(0,t_r)$. That is, the coupling laser is not
effected very much by the resonant medium and it propagates at speed $c$\ without distortion.
Thus, in this limit we only need to solve Eq.(11a) which becomes
\begin{equation}
\frac{\partial}{\partial z}\left(\frac{\Om_{p}^{*}\tau}{|\Om_c(0,t_r)\tau|}\right)=-\frac{\ka_{12}\tau}{|\Om_{c}(0,t_r)\tau|^2}\frac{\partial}{\partial t_r/\tau}\left(\frac{\Om_{p}^{*}\tau}{|\Om_{c}(0,t_r)\tau|}\right).
\end{equation}
If we make the following dimensionless substitutions
\begin{subequations}
\begin{eqnarray}
W_p&=&\frac{\Om_p^{*}}{|\Om_c(0,t_r)|},\\
v(t_r)&=&\int_{-\infty}^{t_r/\tau}\left(|\Om_{c}(0,t_r^{'})\tau|^{2}\right)d\left(\frac{t_r^{'}}{\tau}\right),\\
u(z)&=&\int_{0}^z\ka_{12}\tau dz^{'}, 
\end{eqnarray}
\end{subequations}
we then find that the solution for $W_p$\ is
\begin{equation}
W_p=F_p(v-u),
\end{equation}
with the function $F_p$\ being determined by using $z=0$ at the entrance of the cell,
\begin{equation}
F_p(v(t))=\frac{\Om_p(0,t)}{|\Om_{c}(0,t)|},
\end{equation}
where $v(t)$\ is given in Eq.(20b), with $t_r=t$\ at $z=0$.
A table of $F_p$\ as a function of its argument can be made, which will be used
when $z$ is no longer zero and the argument is $v(t_r)-u(z)$.  Once we have
tabulated $F_p$, the probe field at different $z$ can be trivially calculated
using the known coupling field, e.g.
\begin{equation}
\Om_{p}^{*}(z,t_r)=|\Om_{c}(0,t_r)|F_p(v(t_r)-u(z)).
\end{equation}
Again, the tabulation of $F_p$\ is determined completely at all later times
by Eq (22) - at least for the case where only a time delayed coupling laser pulse enters
the material at $z=0$.

\section{Adiabatic approximation applied to optical field regeneration}

In the precious sections, we have provided detailed descriptions and examples
on how adiabatic approximation works.  In this section, we will discuss various
aspects of the adiabatic approximation, especially the validity in the context
of probe pulse ``revival".  These discussions provide many subtle understandings on
the process which, in our view, are not discussed either in the recent works
on ``probe pulse revival" and the original studies of adiabatic approximation.  This
section, therefore, constitutes the main contributions of the present study.
\vskip 5pt
The validity of the adiabatic approximation depends on more than just the slow time
dependence of the amplitudes of the probe and coupling lasers.  For instance,
the approximation will fail in dealing with the solution for the atomic response
if the two pulses are sent simultaneously and the pulse widths and shapes
are the same, even with both fields being turned on slowly.  In this situation, at early times substantial populations will be
promoted to state $|2>$, in contradiction to the adiabatic solution which, to
the lowest order, predicts a zero population for this state.  What is needed is to
have the coupling laser built up and establish laser induced transparency before
the probe laser starts to become intense.  This could be done by either time
delaying the probe laser relative to the coupling laser, or by making the
coupling laser pulse length much longer than that of the probe laser and by
having them peak in intensity at the same time.  In the following treatment
we will consider the latter situation.  We therefore will choose the
characteristics of $\Om_{p}(0,t)$\ and $\Om_{c}(0,t)$\ that would
be conducive to making the adiabatic approximation work well, and demonstrate that
this is indeed the case.  We emphasize that our choice of the form of the laser
fields is only one of many workable assumptions on the acceptable forms of the
laser pulses that will make the adiabatic theory work well.
\vskip 5pt
Let us consider the functional forms for $\Om_{p}(0,t)$\ and $\Om_{c}(0,t)$\
as
\begin{subequations}
\begin{eqnarray}
\Om_{p}(0,t)&=&\Om_{p0}e^{-(t/\tau)^2}, \\
\Om_{c}(0,t)&=&\Om_{c0}\left(e^{-0.2(t/\tau)^2}+Re^{-0.2(t/\tau-x_0)^2}\right). 
\end{eqnarray}
\end{subequations}
With this functional form, the  Rabi frequency due to the coupling laser rises up to produce
laser induced transparency before the probe laser becomes large enough to have any effect,
an important consequence of having a significantly longer pulse length for the coupling laser.
Without this characteristic of the coupling laser, the probe laser would produce a population
in state $|2>$\ before the transparency was induced.  The production of population in $|2>$\
would mean that different adiabatic states have already been mixed and we have a failure of
a strictly adiabatic picture in which only a single adiabatic state persists throughout the
pulse.  In Eq.(24), $\Om_{p0}$\ and $\Om_{c0}$\ are real constants characterising the
peak amplitude of the two half-Rabi frequencies before the pulses enter the resonant
medium.  The parameter $\tau$\ is a measure of the pulse length of the probe laser,
$R$\ is the ratio of the Rabi frequency at which the coupling laser recurs to its initial
amplitude, $x_0=t_d/\tau$\ is the value of $t_r/\tau$\ at which the peak of the coupling laser recurs.
\vskip 5pt
We will begin by pointing out how the adiabatic approximation can be used to understand
what happens for $R=0$. That is, for the case where there is no recurring coupling laser pulse.
We take $\Om_{p0}\tau=\Om_{c0}\tau=20$, $\kappa_{12}\tau=\kappa_{32}\tau=700 cm^{-1}$, and $\ga_{2}\tau=0$.
We will soon see that in this numerical example the group velocity is
sufficiently small so that the coupling laser dies away before the probe pulse can propagate through the vapor
cell.  When the coupling laser begins to die out, the intensity of the two lasers becomes
proportional to each other.  This is because when the coupling laser is beginning to become
small, the following relation is appropriate (see Eq.(16c))
\begin{equation}
v(t_r)\simeq |\Omega_{c0}\tau|^2\sqrt{5\pi/2}+|\Omega_{p0}\tau|^2\sqrt{\pi/2}.
\end{equation}
That is, almost all of the area under the two Gaussian pulse shapes has already been included
in the integration. Also, $|\Omega_{p}(0,t_r)|^2$\ has long been very small compared with $|\Om_{c}(0,t_r)|^2$.
Thus, for $t_r/\tau>3.0$\ we have, as a very good approximation
\begin{equation}
\Om_{p}^{*}(z,t_r)\simeq \sqrt{|\Om_{c}(0,t_r)|^2+|\Om_{p}(0,t_r)|^2}F_p((|\Om_{c0}\tau|^2\sqrt{5\pi/2}+|\Om_{p0}\tau|^2\sqrt{\pi/2})-\ka_{12}\tau z).
\end{equation}
(In the case of an inhomogeneous medium $\ka_{12}\tau z$\ would be replaced by
an integral over the $z$ dependent $\ka_{12}\tau$.)  At such a late time during the pulse, since the argument of $F_p$\ depends
only on the $z$ coordinate, therefore the time dependence of
$\Om_{p}^{*}$\ is obviously exactly the same as that of
$\sqrt{|\Om_{c}(0,t_r)\tau|^2+|\Om_{p}(0,t_r)\tau|^2}$.  When $|\Om_c(0,t_r)|$\
is several times larger than $|\Om_p(0,t_r)|$, as always is at such
late times at $z=0$, this means that $\Om_{p}^{*}(z,t_r)$\ has the same time dependence
as $\Om_{c}^{*}(0,t_r)$\ at such late times.  The fact that the two pulses have
the same time dependence as they begin to become very weak is the reason that
populations are left in each of the states $|1>$\ and $|3>$.  If one of the
laser fields died out much more rapidly than the other there would only be population
left in only one state.
\vskip 5pt
>From Eq.(26) we see that if we choose $z$ such that
\begin{equation}
\ka_{12}\tau z=\frac{1}{2}\left(|\Om_{c0}\tau|^2\sqrt{5\pi/2}+\sqrt{\pi/2}|\Om_{p0}\tau|^2\right),\nonumber
\end{equation}
then the argument of $F_p$\ will be the same as at $z=0$\ and $t_r=t=0$.  With this
argument, the value of $F_p$\ is $F_p=\Om_{p0}/\sqrt{|\Om_{c0}|^2+|\Om_{p0}|^2}$.
This is the largest value that $F_p$\ takes on.  Thus, at this depth into
the medium the value of $|A_3(z,t_r)|$\ matches its largest value at $z=0$.
This population persists at large $t_r$\ until very slow collisional effects either
destroy the coherence left behind in states $|1>$\ and $|3>$, or until
collisions mix the populations of the two states.  This long persistence of
a coherent mix of populations in states $|1>$\ and $|3>$\ is what leads to the
``revival" of an optical field with frequency very close to that of the original
probe laser when a coupling laser is re-injected into the medium
after some time delay [11]. The number of atoms left in $|3>$\ is (within the
adiabatic approximation)
equal to the number of photons in the original probe pulse (recall that
we have chosen parameters so that the probe pulse does not penetrate the
entire medium).  The creation of a coherent mixed stat is also what makes
this situation interesting from the point of view of quantum computing.
\vskip 5pt
Next, we investigate the situation where $R>0$, i.e. a second coupling pulse enters
the medium at a delayed time.  Let
$\Om_{c0}\tau=20$, $\Om_{p0}\tau=5$, $\ga_2\tau=0$,
$\kappa_{12}\tau=\ka_{32}\tau=200 cm^{-1}$.  From the above consideration,
we expect $|A_3|$\ to have its largest value at large $t_r$\ when $z=2.86$\ cm.
In Figure 2 we show a contour plot of $|A_3(z,t_r)|$\ based on our adiabatic
theory.  The predicted $|A_3|$ is indicated
by the line of constant color leading from $t_r=0$\ and $z=0$\ out to the
horizontal path at $z\simeq 2.86$\ cm, as expected.
In Figure 3 we show a surface plot of the probe field Rabi frequency
$\Omega_{p}\tau$\ as functions of $t_r/\tau$\
at $z=3$\ cm.  The long asymmetric tail of $\Omega_p$\ is a consequence of the behavior described
in Eq.(26). In the region between $2.5\le{t_r}/\tau\le{5}$ the ratio of the two half-Rabi frequencies
is close to constant, averaging around 0.24.  This ratio is also close to the
adiabatic approximation for $A_3$ since $|\Om_{p}|^2<<|\Om_{c}|^2$. In a different set of parameters,
we choose $\Om_{c0}\tau=\Om_{p0}\tau=20$,
$\ga_{2}\tau=0$, $R=2.923$, and
$\kappa_{12}\tau=\kappa_{32}\tau=700 cm^{-1}$.  In this example,our theory predicts
that the
peak value of $|A_3(t_r,z)|$\ before the second coupling laser pulse enters the medium
will occur at $z=1.159$ cm. Indeed, as can be seen clearly from Figure 4 that
$|A_3(z,t_r)|$\ takes on the value $1/\sqrt{2}$\ at a depth of about 1.1-1.2 cm
into the medium, as predicted.  Figure 5 depicts the corresponding
probe field predicted with adiabatic theory for this set of parameters.
\vskip 5pt
We now illustrate how one can take a set of initial conditions and
very simply determine the characteristics of the ``revived" probe pulse at the
point it emerges from the atomic vapor.

\vskip 5pt
First, note that with our choice of
pulse characteristics and during the recurring coupling laser pulse
\begin{subequations}
\begin{eqnarray}
v(t_r)&=&S+\frac{R^2}{2}|\Om_{c0}\tau|^2\sqrt{\frac{5\pi}{2}}\left(1+erf\left(\sqrt{\frac{2}{5}}\frac{(t_r-t_d)}{\tau}\right)\right),\\
S&=&|\Om_{c0}\tau|^2\sqrt{5\pi/2}+|\Om_{p0}\tau|^2\sqrt{\pi/2}.  
\end{eqnarray}
\end{subequations}
The reader will recall that $F_p$\ was determined from the functional dependence
of the probe and coupling laser pulses at $z=0$.  In particular, $W_p(v(t))=-A_3(0,t)$,
where $A_3(0,t)$\ is determined from Eq.(6c) and the pulse characteristics at $z=0$.
In addition, the following properties of the function $F_p$ are very useful:
\begin{subequations}
\begin{eqnarray}
F_p(0)&=&0, \\
F_p(S)&=&0, \\
F_p(S/2)&=&F_p(v(0))=\frac{\Om_{p0}\tau}{\sqrt{|\Om_{c0}\tau|^2+|\Om_{p0}\tau|^2}}.
\end{eqnarray}
\end{subequations}
As will be seen later, we will invoke these properties when $z=z_m$ is reached,
where $z_m$\ is the value of $z$ at the end of the medium, to estimate
the time interval over which the ``revived" probe pulse exits, as well as the
peak amplitude of $\Om_{p}^{*}(z_m,t_r)$.

\vskip 5pt
Before proceeding further, we will
point out some obvious properties that must be satisfied by the parameter, $R$,
that appears in Eq.(24b).  We first note
that $R$\ must be large enough so that $v(t_r)-\ka_{12}\tau z_m>0$.  That is, $R$ must
be large enough so that the group velocity becomes large enough to allow most
of the regenerated photons to exit the cell before the laser
induced transparency ends. This group velocity depends on position and time as (see Eq.(15))
\begin{equation}
\frac{c}{v_g}=1+\frac{\ka_{12}c\tau^2}{|\Om_p(z,t_r)\tau|^2+|\Om_c(z,t_r)\ta|^2}.
\end{equation}
\vskip 5pt
We first write down the condition for a time $t_{r1}$ at which the ``revived probe pulse" first
reaches $z_m$.  This is the earliest time at which the argument of $F_p$\ goes
from being negative and becomes zero. At this point $F_p(0)=0$, but at later times
it will become non-zero. This time is determined by
\begin{equation}
v(t_{r1})-\ka_{12}\tau z_m=0.
\end{equation}
At time $t_{rm}$ the value of $F_p(S/2)$\ will be equal to $-A_3(0,0)$.  This is
the maximum value $F_p$\ takes on with the set of parameters chosen.  This time
is determined by
\begin{equation}
v(t_{rm})-\ka_{12}\tau z_m=S/2.
\end{equation}
Finally, at the time $t_{r2}$ at which the ``revived probe pulse" completes
its exit from the medium.  At this time, we have $F_p(S)=0$.  Therefore,
\begin{equation}
v(t_{r2})-\ka_{12}\tau z_m=S.
\end{equation}
Using Eqs.(27a,b) we can rewrite Eqs.(30-32) as
\begin{subequations}
\begin{eqnarray}
erf\left(\sqrt{2/5}\frac{(t_{r1}-t_d)}{\tau}\right)&=&\frac{2(\ka_{12}\tau z_m-S)}{R^2|\Om_{c0}\tau|^2\sqrt{5\pi/2}}-1,\\
erf\left(\sqrt{2/5}\frac{(t_{rm}-t_d)}{\tau}\right)&=&\frac{2(\ka_{12}\tau z_m-S/2)}{R^2|\Om_{c0}\tau|^2\sqrt{5\pi/2}}-1. \\
erf\left(\sqrt{2/5}\frac{(t_{r2}-t_d)}{\tau}\right)&=&\frac{2(\ka_{12}\tau z_m)}{R^2|\Om_{c0}\tau|^2\sqrt{5\pi/2}}-1, 
\end{eqnarray}
\end{subequations}
which can be further condensed into
$$
erf\left(\sqrt{2/5}\frac{(t_r-t_d)}{\tau}\right)=\frac{\al-f\be}{R^2}-1,
$$
where
\begin{eqnarray}
\al&=&\frac{2\ka_{12}\tau z_m}{|\Om_{c0}\tau|^2\sqrt{5\pi/2}},\nonumber\\
\be&=&\frac{2S}{|\Om_{c0}\tau|^2\sqrt{5\pi/2}},\nonumber
\end{eqnarray}
with $f=1$\ to determine $t_{r1}$, $f=0.5$\ to determine $t_{rm}$, and
$f=0$\ to determine $t_{r2}$.  Correspondingly, we obtain the peak value of
$\Om_{p}^{*}(z_m,t_{rm})\tau$\ as
\begin{equation}
\Om_{p}^{*}(z_m,t_{rm})=R\Om_{c0}e^{-(t_{rm}-t_d)^2/(5\tau^2)}\frac{\Om_{p0}\tau}{\sqrt{|\Om_{c0}\tau|^2+|\Om_{p0}\tau|^2}}.
\end{equation}
\vskip 5pt
As an example, let $\ka_{12}\tau=\ka_{32}\tau=200 cm^{-1}$, $z_m=8$\ cm, $\Om_{c0}\tau=20$,
$\Om_{p0}\tau=5$, $R=4$, and $\ga_{2}\tau=0$.  With these parameters, we find
$\al=2.854598$, $\al/R^2=0.1784124$, $\be=2.0559017$, and $\be/R^2=0.128494$.
Thus, $(t_{r1}-t_d)/\tau=-2.19$, $(t_{rm}-t_d)/\tau=-1.767$, and $(t_{r2}-t_d)/\tau=-1.50$.
The value of $\Om_{p}^{*}(z_m,t_{rm})$\ is thus estimated as
$$
\Om_{p}^{*}(z_m,t_{rm})=4(20)\exp(-0.2(1.767)^2)\frac{5}{\sqrt{20^2+5^2}}=10.39.
$$
The time $t_{rm}$ corresponds to the largest value of $F_p$, but not quite to the maximum
of $\Om_{p}^{*}(z_m,t_r)$.  The actual maximum occurs closer to -1.74 with
the maximum value of $10.44$. However, our simple estimate usually does very
well if $R$\ is large enough to easily let all of the regenerated probe laser
photons escape.
\vskip 5pt
We next note that if $R$\ is chosen to make the argument of $F_p$\ exactly the same
as its value at $z=0$ and $t_r=t=0$\ at the time $t_r=t_d$, we then have
\begin{equation}
|\Om_{p0}\tau|^2\sqrt{\frac{\pi}{2}}+|\Om_{c0}\tau|^2\sqrt{\frac{5\pi}{2}}\left(1+\frac{R^2}{2}\right)-\ka_{12}\tau z_m=\frac{1}{2}\left(|\Om_{c0}\tau|^2\sqrt{\frac{5\pi}{2}}+|\Om_{p0}\tau|^2\sqrt{\frac{\pi}{2}}\right).
\end{equation}
Thus,
\begin{equation}
|\Om_{p0}\tau|^2\sqrt{\pi/8}+|\Om_{c0}\tau|^2\sqrt{5\pi/8}(1+R^2)=\ka_{12}\tau z_m.
\end{equation}
By choosing the value of $R$\
such that this equation is satisfied, one achieves a situation where the peak
value of $\Om_{p}(z_m,t_d)$\ is given by (see Eq.(34))
\begin{equation}
\Om_{p0}^{*}(z_m,t_d)=R|\Om_{c0}|\frac{\Om_{p0}\tau}{\sqrt{|\Om_{c0}\tau|^2+|\Om_{p0}\tau|^2}}.
\end{equation}
\vskip 5pt
The characteristics of this regenerated field can be estimated as follows.
We first determine the value of $t_r$\ such that the right hand side of
Eq.(32) is zero or the full area under square of the  incident probe and
coupling lasers.  This yields a range of time over which the probe rises from
zero and returns to zero at the exit of the cell.  One half of this time
interval is close to the
full-width at half-maximum of the exiting probe laser beam.  We then
note that our condition on $R$\ yields
\begin{equation}
R^2=\frac{\ka_{12}\tau z_m-|\Om_{c0}\tau|^2\sqrt{5\pi/8}-|\Om_{p0}\tau|^2\sqrt{\pi/8}}{|\Om_{c0}\tau|^2\sqrt{5\pi/8}}.
\end{equation}
Using these considerations we estimate the full-width at
half-maximum pulse length, in the unit of the original probe pulse length $\tau$, to be
\begin{equation}
\De_{1/2}=\frac{\sqrt{5\pi/2}}{R^2}\left(1+\frac{1}{\sqrt{5}}\left|\frac{\Om_{p0}}{\Om_{c0}}\right|^2\right).
\end{equation}
If the original probe laser has only penetrated a small fraction of the thickness of the
medium when the coupling laser pulse has passed by, then $R$\ will turn out
to be much larger than unity.
This means that  the width of the regenerated field is generally much smaller
than the width of the initial probe pulse.  Correspondingly, the bandwidth of
the light will be larger.  An immediate conclusion is thus available: the regenerated
pulse is not the replica of the original probe pulse.  It is coherent, to an extended
degree, with the
coupling laser because of the stimulated nature of the regeneration process.  Therefore,
the notion of ``coherent storage" of the information carried in the original probe field
is not accurate [7-9]. These features can be seen from Figures 3 and 5 which
show $\Om_{p}^{*}$\ as a function of $t_r/\tau$\ and $z$, as well as the time
dependence of the regenerated field at the exit of the medium.  In particular,
as can be clearly seen from Figures 3 and 5, there is no optical field left
in the region between the time when the original
probe pulse and the companion coupling pulse enter the medium and the time when the
second coupling pulse enters the medium.  Therefore, photons are neither
``stored" nor ``stopped".  All photons from the original probe pulse are converted to the
population of the state $|3>$.  These figures also show that in
``reviving the probe pulse" every regenerated photon
comes at the expense of flipping population from state $|3>$\ to state
$|1>$. When the population of $|3>$\ has been exhausted, there can be no further
photon generated.  Thus, in cases where the exiting regenerated field
has a peak intensity
much larger than the initial probe beam, the pulse width of this exiting pulse
must be necessarily and correspondingly narrower.  This is why the peak intensity of
the regenerated field can go as $R^2$, while the width of the pulse tends to
go as $1/R^2$.

\section{Comparison between numerics and the adiabatic theory}

We will now present some examples showing the degree of agreement between
numerical calculations and the adiabatic theory developed in the previous sections.
To obtain the numerical
results, we solve numerically the coupled Maxwell and time dependent
Schr\"{o}dinger equations
\begin{subequations}
\begin{eqnarray}
\left(\pd{A_1}{t_r}\right)_z&=&i\Om_{p}A_2, \\
\left(\pd{A_2}{t_r}\right)_z&=&i\Om_{p}^{*}A_1+i\Om_{c}^{*}A_3+i\left(\delta + i\frac{\ga_2}{2}\right)A_2,\\
\left(\pd{A_3}{t_r}\right)_z&=&i\Om_{c}A_2,\\
\end{eqnarray}
\end{subequations}

\begin{subequations}
\begin{eqnarray}
\left(\pd{\Om_{p}^{*}}{z}\right)_{t_r}&=&i\ka_{12}A_1^{*}A_2,\\
\left(\pd{\Om_{c}^{*}}{z}\right)_{t_r}&=&i\ka_{32}A_3^{*}A_2.
\end{eqnarray}
\end{subequations}
Here, $A_1,A_2,$\ and $A_3$\ are the same as in Eqs.(2a-2c) except for the
removal of the $z$ dependent phase factors.  First, recall the example that we
have studied before where
$\Om_{c0}\tau=20$, $\Om_{p0}\tau=5$, $\ga_2\tau=0$,
$\kappa_{12}\tau=\ka_{32}\tau=200 cm^{-1}$.  The discussions based on the
adiabatic approximation predicts that a population of
$|A_3|=1/\sqrt{17}=0.242536$ at a depth of $z=2.86$\ for $t_r/\tau>6.0$,
as can be seen from Fig. 2.  As a comparison, we show, in Figure 6,
a contour plot of $|A_3(z,t_r)|$ for the same parameters given
in Figure 2 but using numerical solution of Eqs.(40-41).  The flat region in
the middle, a region where both of the probe and coupling lasers have been
extinguished, indicates that during this period of time the population in
the two-photon state remains nearly constant, just as we have concluded before
based on the adiabatic theory that
a population is left behind in states $|3>$\ and $|1>$\ which persists for
a long time.  Correspondingly, we show,
in Figure 7, a surface plot of the probe field resulted from the direct
numerical solution of Eqs.(40-41) using the same parameters as in Figure 3.
We note that there is no probe laser field between
$t_r/\tau=2$\ and the time of
the injection of the second coupling laser pulse, the same period where the
population in the two-photon state remains nearly constant.  This indicates,
as our adiabatic theory has shown,
that photons are neither ``stopped" or ``stored" in the
medium as claimed in Ref. [7-9].  As the original probe photons
being slowed down, they are absorbed to produce the two-photon excitation
with the above noted population in the state $|3>$.  When all probe photons
have been absorbed, there will be no further increase of the population in
state $3>$.  One may still argue that the probe photons are ``stored" in the form
of atomic coherence involving the two-photon excitation, and later when a second coupling
pulse is injected into the medium, the probe pulse is ``retrived" from this persist
atomic coherence.  This is, however, not correct.  As we have shown the regenerated
pulse has very different time-spectra characteristics from that of the
original probe pulse.  Therefore, the concept of ``retriving the probe pulse"
is incorrect.  Indeed, the Fourier transform spectrum of all the regenerated
fields have shown both different frequency content and noise characteristics that
were not there in the original probe pulse.  A simple argument based on our
result (see Eqs.(37,39)) is sufficient to show that the regenerated pulse is very different
from the original probe pulse, that is the regenerated pulse contains information
in the second coupling pulse.  Therefore, can not be the faithful replica of the
original probe pulse alone.

\vskip 10pt
\noindent Further comparison between the adiabatic theory and the numerical solution
are presented in Figures 8 and 9 for the second set of parameters used to create
Figures 4 and 5 which were obtained with adiabatic theory.  Finally,
we show a case where non-vanishing one-photon detuning is assumed.  Figure 10
is surface plot of the probe field obtained from the adiabatic theory with $\de\tau=10$, whereas Figure 11 is obtained
by directly solving Eqs.(40-41) using the same parameters.  All these comparisons
show very good agreement between our theory and numerical solutions.
\vskip 5pt
The numerical calculations were also tested by making use of the fact that with $\ga_2\tau=0$,
the sum of the squares of the three state amplitudes should be unity.  In all numerical
examples described in this work, this condition of unity was preserved through at least seven significant figures if $\ga_2\tau=0$.
Also, in all cases when the number of $t_r/\tau$\ and $z$ grid points were doubles
for all figures shown, the values of the state amplitudes and half-Rabi frequencies repeated through
five significant figures at all points where the functions were not near a zero.

\section{Case of far from resonance}

In this Section, we examine the far off resonance excitation, a case has never
been studied in literature with the same adiabatic treatment described in previous sections.
Specifically, we investigate the condition for $\la_0\tau$\ to stay well
separated from $\la_{\pm}\tau$. We will show analytically that the formulism
for far detuned excitation exhibits detuning independent feature when a specific
condition is satisfied, a surprising result that to the best of our knowledge
has never been reported in the literature.
\vskip 5pt
It is easy to see the condition for the eigenvalues to be well separated when $|\de|>>|\Om_c|$\ and
$|\de|>>|\Om_p|$.  In this limit
\begin{subequations}
\begin{eqnarray}
\sqrt{|\Om_p|^2+|\Om_c|^2+\left(\frac{\de+i\ga_2/2}{2}\right)^2}&&\simeq{\frac{\de+i\ga_2/2}{2}}\left(1+2\frac{|\Om_p|^2+|\Om_c|^2}{(\de+i\ga_2/2)^2}\right)\nonumber\\
&=&\left(\frac{\de+i\ga_2/2}{2}\right)+\frac{|\Om_c|^2+|\Om_p|^2}{\de+i\ga_2/2}.\nonumber
\end{eqnarray}
\end{subequations}
The key element that could invalidate the adiabatic treatment is the eigenvalue
$$
\la_{-}=-\frac{|\Om_c|^2+|\Om_p|^2}{\de+i\ga_2/2}\simeq\frac{|\Om_{c}|^2+|\Om_{p}|^2}{\de},
$$
which could become very close to the eigenvalue of the dark state $lamda_0=0$
at sufficiently large detuning.  Since the length of the pulse due to the coupling field is chosen to be much longer than that due to the
probe laser, this eigenvalue can only be well separated from the zero eigenvalue at times when the probe laser
first starts to build up if $|\la_{-}\tau|>>1$\ at such times.  Physically, what is required is for the ac Stark
shift in level $|3>$\ to satisfy
$$
\frac{|\Om_c|^2}{\de}\tau >> 1.
$$
This means that as the two-photon transition between $|1>$\ and $|3>$\ is
driven, the final state, i.e. state $|3>$, must be shifted
out of exact resonance by an amount that is much larger than the laser
bandwidth.  (We have assumed that the natural
width of level $|3>$ is very small, being due to collisional relaxation in a
very cold vapor.  If the laser bandwidth is less than the width of the state $|3>$,
the latter will replace $\tau ^{-1}$ in the above equation.)  In Figure 12,
we first show a surface plot of $\Om_p(z,t)$
based on our adiabatic theory for zero detuning.
Figure 13 is a corresponding plot where the full numerical
evaluations of Eqs.(40-41) are carried out.  As expected, the solution based
on adiabatic theory agrees well with the numerical solution.  We now compare
Figures 12 and 13 with Figures 14 and 15 which are the surface plots of
$\Om_p(z,t)$ with large one-photon detuning $\de\tau=120$.  Figure 14
is based on our adiabatic theory, therefore, should be compared with Figure 12, whereas
Figure 15 is the result of full numerical calculation, thus should be compared
with Figure 13.  From these figures, we therefore conclude that when the appropriate adiabatic
condition is met, the solution to the problem is insensitive to the detuning.
Indeed, when the conditions given above
are satisfied, the adiabatic approximation will be valid for all smaller $\de$ as well,
a surprising results that has not been reported in literature and could play
an important role in the future experiments.  

\section{Conclusion}

In conclusion, we have shown in detail how the adiabatic approximation can be used
to understand the propagation of a pair of optical pulses in a highly dispersive
medium.  Starting from the adiabatic solution of a three-level system interacts
with two laser fields, we showed analytically that in the case where the one-photon detuning
is small (i.e. $0\le\de<|\Om_c|$), a coherent optical
field with the frequency very close to that of the original probe can be regenerated
when a coupling laser pulse in re-injected into the medium at a delayed time.
We have provided detailed analysis
on the conditions and characteristics of the regenerated field, including the
estimate of the pulse width.  Our analysis has shown that the regenerated pulse
has a very different width therefore cannot be a perfect replica of the original
probe pulse.  Therefore, the notion of ``coherent storage of the original probe
pulse" is an incorrect description.  As expected, the regenerated pulse is
coherent, to an extended degree, with the coupling laser but not the original
probe field.  Subsequent
numerical calculations have confirmed these analytical results.
We have further developed the adiabatic theory to the cases where
the one-photon detuning is larger enough so that $|\Om_c|<\de$ and obtained
a different type of adiabatic condition and solution.  For time varying probe and
coupling fields, the highly nonlinear wave equation for the recurring probe field
is then investigated numerically. It is shown that even with this large one-photon
detuning, as long as one chooses the pulse envelope and sequence properly so
that there is a coherence left in the system, then there will be a coherent
optical field generated when the coupling laser is turned on only.

\vskip 15pt
\leftline{We thank Professor R. Grobe for providing reprints.}

\section*{List of References}

\begin{itemize}
\item[1.]{D. Grischkowsky, M.M.T. Loy, and P.F. Liao, Phys. Rev. A12, 2514 (1975).}
\item[2.]{J. Oreg, F.T. Hioe and J.H. Eberly, Phys. Rev A29,690 (1984); J.R. Kuklinski,
U. Gaubatz, F.T. Hioe, and K. Bergmann, Phys. Rev. A 40 R6749 (1989).}
\item[3.]{K.J. Boller, A. Imamoglu, and S.E. Harris, Phys. Rev. Lett. 66, 2593 (1991).}
\item[4.]{R. Grobe and J.H. Eberly, Laser Physics 5, 542 (1995); J.H. Eberly,
A. Rahman, and R. Grobe, Laser Physics 6, 69 (1996).}
\item[5.]{J.R. Csesznegi and R. Grobe, Phys. Rev. Lett. 79, 3162 (1997); \
J.R. Csesznegi, B.K. Clark, and R. Grobe, Phys. Rev. A57, 4860 (1998); R. Grobe,
Acta Physica Polica A93, 87 (1998).}
\item[6.]{L.V. Hau et al., Nature (London) 397,594 (1999); M.M. Kash et al.,
Phys. Rev. Lett. 82, 5229 (1999); D.Budker et al., Phys. Rev. Lett. 83, 1767
(1999).  For early works on the reduction of the group
velocity of optical field in resonant media, see O.Schmidt, R. Wynands,
Z. Hussein, and D. Meschede, Phys. Rev. A53, R27(1996); G. Muller, A. Wicht, R. Rinkleff,
and K. Danzmann, Opt. Communi. 127, 37 (1996); A.M. Akulshin, S. Barreiro, and
A. Lezama, Phys. Rev. Lett. 83, 4277 (1999).}
\item[7.]{C. Liu et al., Nature (London) 409, 490 (2001).}
\item[8.]{D.F. Phillips et al., Phys. Rev. Lett. 86, 783 (2001).}
\item[9.]{E.A. Cornell, Nature (London), 409, 461 (2001).}
\item[10.]{M. Fleischhauer, Opt. Express 4, 107 (1999); M. Fleischhauer, S.F. Yelin,
and M.D. Lukin, Opt. Communi. 179, 395 (1999); M. Fleischhauer and M.D. Lukin,
Phys. Rev. Lett. 84, 5094 (2000).}
\item[11.]{Such a regeneration of an optical field is not unique to the process
described in [7-9].  In fact, it is just a stimulated Raman generation in the context of
electromagnetically induced transparency.  It is not surprising at all that the
frequency of the regenerated field is close to that of the original probe field
because the second coupling pulse is identical to the first one (in frequency).
The regenerated field, however, is not the faithful copy of the probe field as
we have shown later.  Therefore, the statement of ``reviving the probe pulse" in Refs.[7-9]
is inaccurate.  In fact, one can mix the two lower states by pulsing a magnetic field,
and then turning on a coupling laser to regenerate a very
similar optical field.}
\end{itemize}

\section*{Figure Captions}

\begin{itemize}

\item[Figure 1.]{Energy level diagram showing relevant laser excitations.  The
decay rate of the state $|3>$ is assumed to be very small and hance neglected.}

\item[Figure 2.]{Contour plot of $A_3(z,t)$ based on adiabatic theory.  Parameters
used: $\Om_p\tau=5$, $\Om_c\tau=20$, $\ga_2\tau=0$, $\kappa_{12}\tau=\kappa_{32}\tau=200 cm^{-1}$,
$R=4$, $t_d/\tau=11$.}

\item[Figure 3.]{Surface plot of $\Om_p(z,t)$ based on adiabatic theory. Parameters
used: $\Om_p\tau=5$, $\Om_c\tau=20$, $\ga_2\tau=0$, $\kappa_{12}\tau=\kappa_{32}\tau=200 cm^{-1}$,
$R=4$, $t_d/\tau=11$.}

\item[Figure 4.]{Contour plot of $A_3(z,t)$ based on adiabatic theory.  Parameters
used: $\Om_p\tau=\Om_c\tau=20$, $\ga_2\tau=0$, $\kappa_{12}\tau=\kappa_{32}\tau=700 cm^{-1}$,
$R=3$, $t_d/\tau=11$.}

\item[Figure 5.]{Surface plot of $\Om_p(z,t)$ based on adiabatic theory.  Parameters
used: $\Om_p\tau=\Om_c\tau=20$, $\ga_2\tau=0$, $\kappa_{12}\tau=\kappa_{32}\tau=700 cm^{-1}$,
$R=3$, $t_d/\tau=11$.}

\item[Figure 6.]{Contour plot of $A_3(z,t)$ based on numerical solution of Eqs.(40-41).
Parameters used: $\Om_p\tau=5$, $\Om_c\tau=20$, $\ga_2\tau=0$,
$\kappa_{12}\tau=\kappa_{32}\tau=200 cm^{-1}$, $R=4$, $t_d/\tau=11$.  This figure
should be compared with the Figure 2.}

\item[Figure 7.]{Surface plot of $\Om_p(z,t)$ based on numerical solution of Eqs.(40-41).
Parameters used: $\Om_p\tau=5$, $\Om_c\tau=20$, $\ga_2\tau=0$,
$\kappa_{12}\tau=\kappa_{32}\tau=200 cm^{-1}$, $R=4$, $t_d/\tau=11$. This figure
should be compared to Figure 3.}

\item[Figure 8.]{Contour plot of $A_3(z,t)$ based on numerical solution of Eqs.(40-41).
Parameters used: $\Om_p\tau=\Om_c\tau=20$, $\ga_2\tau=0$,
$\kappa_{12}\tau=\kappa_{32}\tau=700 cm^{-1}$, $R=3$, $t_d/\tau=11$. This figure
should be compared with Figure 4.}

\item[Figure 9.]{Surface plot of $\Om_p(z,t)$ based on numerical solution of Eqs.(40-41).
Parameters used: $\Om_p\tau=\Om_c\tau=20$, $\ga_2\tau=0$,
$\kappa_{12}\tau=\kappa_{32}\tau=700 cm^{-1}$, $R=3$, $t_d/\tau=11$.  This figure
should be compared with Figure 5.}

\item[Figure 10.]{Surface plot of $\Om_p(z,t)$ based on adiabatic theory with
non-vanishing one-photon detuning.  Parameters used: $\Om_p\tau=10$, $\Om_c\tau=20$,
$\ga_2\tau=0$, $\de\tau=10$, $\kappa_{12}\tau=\kappa_{32}\tau=200 cm^{-1}$, $R=4$,
$t_d/\tau=11$.}

\item[Figure 11.]{Surface plot of $\Om_p(z,t)$ based on numerical solution of
Eqs.(40-41) with non-vanishing one-photon detuning.  Parameters used:
$\Om_p\tau=10$, $\Om_c\tau=20$, $\ga_2\tau=0$, $\de\tau=10$,
$\kappa_{12}\tau=\kappa_{32}\tau=200 cm^{-1}$, $R=4$, $t_d/\tau=11$.}

\item[Figure 12.]{Surface plot of $\Om_p(z,t)$ based on adiabatic
theory with zero detuning.  Parameters used: $\Om_p\tau=10$, $\Om_c\tau=40$,
$\ga_2\tau=0$, $\de\tau=0$, $\kappa_{12}\tau=\kappa_{32}\tau=200 cm^{-1}$, $R=4$,
$t_d/\tau=10$.}

\item[Figure 13.]{Surface plot of $\Om_p(z,t)$ based on
numerical evaluation of Eq.s(40-41) with zero detuning.
Parameters used: $\Om_p\tau=10$, $\Om_c\tau=40$,
$\ga_2\tau=0$, $\de\tau=0$, $\kappa_{12}\tau=\kappa_{32}\tau=200 cm^{-1}$, $R=4$, $t_d/\tau=10$.
This figure should be compared with Figure 12.}

\item[Figure 14.]{Surface plot of $\Om_p(z,t)$ based on adiabatic
theory with large one-photon detuning.  Parameters used: $\Om_p\tau=10$, $\Om_c\tau=40$,
$\ga_2\tau=0$, $\de\tau=1200$, $\kappa_{12}\tau=\kappa_{32}\tau=200 cm^{-1}$,
$R=4$, $t_d/\tau=10$.  This figure should be compared with Figure 12.}

\item[Figure 15.]{Surface plot of $\Om_p(z,t)$ based on
numerical evaluation of Eq.s(40-41) with large one-photon detuning.
Parameters used: $\Om_p\tau=10$, $\Om_c\tau=40$,
$\ga_2\tau=0$, $\de\tau=120$, $\kappa_{12}\tau=\kappa_{32}\tau=200 cm^{-1}$, $R=4$, $t_d/\tau=10$.
This figure should be compared with Figure 13.}
\end{itemize}
\end{document}